\documentclass[epj]{svjour}
\usepackage{graphics}
\usepackage{mathtext}
\usepackage[pdftex]{graphicx}
\newcommand{\eqref}[1]{(\ref{#1})}
\newcommand{\el}{{\rm el}}

\begin{document}
\title{Collision of viscoelastic bodies: Rigorous derivation of dissipative force}
\author{Denis S.\ Goldobin~\inst{1,2,3} \and Eugeniy A.\ Susloparov~\inst{2} \and Anastasiya V.\ Pimenova~\inst{1} \and Nikolai V. Brilliantov \inst{3}}
\institute{
 Institute of Continuous Media Mechanics, UB RAS, Perm 614013, Russia \and
 Department of Theoretical Physics, Perm State University, Perm 614990, Russia \and
 Department of Mathematics, University of Leicester, Leicester LE1 7RH, UK }


\abstract{We report a new theory of dissipative forces acting between colliding viscoelastic bodies. The impact velocity is assumed not to be large, to avoid plastic deformations and fragmentation at the impact. The bodies may be of an arbitrary convex shape and of different materials. We develop a mathematically rigorous perturbation scheme to solve the continuum mechanics equation that deals with both displacement and displacement rate fields and accounts for the dissipation in the bulk of the material. The perturbative solution of this equation allows to go beyond the previously used quasi-static approximation and obtain the dissipative force. This force does not suffer from the physical inconsistencies of the  latter approximation and depends on particle deformation and deformation rate. 
}

\maketitle

\section{Introduction}
Granular materials are abundant in nature; they range from sand and powders on Earth to planetary rings and  dust clouds in outer space \cite{DryGranMed,GranRev,PhysGranMed,SanPowGr,PlanRings}. These material exhibit very unusual properties, demonstrating solid-like, liquid-like or gas-like \cite{book,GranGas1,GranGas2,GasRev} behavior depending on the external load or  magnitude of agitation \cite{wildman,menon,rotdriv}. The physical reason for many unusual phenomena in granular media is the nature of inter-particles interactions in these systems. Contrary to molecular or atomic systems, where particles interact only trough conservative, elastic forces, the interaction between granular particles include dissipative forces. This happens because the grains are themselves macroscopic bodies, which contain macroscopically large number of microscopic degrees of freedom. Hence, during an impact of such bodies their  mechanical energy, associated with the translational or rotational motion, or with the elastic deformation of the particles, is partly transformed into the internal degrees of freedom, that is, into heat. In many applications however, the temperature increase of the grains is insignificant and may be neglected \cite{book}. Obviously, for an adequate description of granular media one needs a quantitative model of inter-particles forces, which includes both elastic and dissipative interactions.

The elastic part of the inter-particle force is known for more than a century from the famous work of Hetrz~\cite{Hertz:1882}. Hertz obtained a mathematically rigorous result for the force acting between elastic bodies at a contact, provided the deformation of the bodies is small as compared to their size; the theory has been developed for particles of an arbitrary convex shape. In spite of a large importance for applications, the rigorous derivation of the dissipative force is still lacking.  Presently there exist phenomenological expressions for the dissipative force, which exploit either linear, e.g.~\cite{PoeschelSchwager2005,LudingReview2008} or quadratic~\cite{Poschl1928} dependence of the force on the deformation rate. Neither linear, nor quadratic dependence, however, is consistent with the experimental data, e.g.~\cite{PoeschelSchwager2005,Poeschel2011}. A derivation of the dissipative force from the first-principles has been undertaken in Ref.~\cite{Pao1955}. A very restrictive approximation used in this work -- the assumption that only the shear deformation is important, substantially undermines its application. A complete derivation of the dissipative force between viscoelastic bodies from the continuum mechanics equations has been done only recently~\cite{bshp96} within a \emph{quasi-static} approximation. In this approximation it is assumed that the displacement field in the bulk of colliding bodies completely coincides with that for a static contact~\cite{bshp96}. The correct functional dependence of the dissipative force, derived in Ref.~\cite{bshp96} has been already suggested (without any rigorous mathematic analysis) in the earlier work of Kuwabara and Kono~\cite{kk87}. In the later studies~\cite{Zheng1,Zheng2} a flaw in the  derivation of the dissipative force of  Ref.~\cite{bshp96} has been corrected. Still the restrictive assumption of the quasi-static approximation has been exploited~\cite{Zheng1,Zheng2}.

Physically, the quasi-static approximation assumes the immediate response of the particle' material to the external load. Two conditions are to be fulfilled in order to make this approximation valid: (i) the characteristic deformation rate should be much smaller than the speed of sound in the system and (ii) microscopic relaxation time of the particle's material should be much shorter than the duration of the impact. The microscopic relaxation time quantifies the time needed for the material of a deformed body to respond to the applied load; in what follows we will give the detailed definition of this quantity. To go beyond the quasi-static approximation, that is, to take into account the deviation of the displacement field in the bulk of a deformed body from the static displacement field, we develop a perturbation approach based on small parameter -- the ratio of microscopic relaxation time and collision duration. In the most of important applications this ratio is indeed small, which implies that for the first time we rigorously derive a dissipative force acting between viscoelastic particles. Although the quasi-static approximation is based on the physically plausible approach, it possesses some inconsistency. This inconsistency is not so visible for a collision of particles of the same material. At the same time when particles of different materials suffer an impact,  the quasi-static approximation predicts non-equal dissipative forces acting between the bodies, which definitely violates the third Newton's law. The other inconsistency is related to the dependence of the dissipative force on the Poisson ratio -- within the quasi-static approximation one obtains zero dissipative force for the case of vanishingly small elastic shear module; this is definitely not physical. These difficulties of the quasi-static approximation are discussed in detail below.

Our new theory, based on the perturbation scheme, is mathematically rigorous and the obtained dissipative force is free from the above inconsistencies. While in the present work we analyze a general case of an impact of viscoelastic bodies of an arbitrary shape and of different materials, the results for a  more simple case of a collision of a sphere with un-deformable plane has been reported earlier \cite{BrillEPLCont2014}.

The rest of the paper is organized as follows. In the next Sec.~II we introduce the equation of motion of viscoelastic medium which we  solve for the case of interest in the next sections. In Sec.~III the solution for the static contact is considered; here we illustrate the general approach and derive the classical Hertz law. In Sec.~IV the dynamic contact is addressed. We elaborate the perturbation scheme and using this scheme derive in Sec.~V the next-order solution. In Sec.~VI we present our new theory for the dissipative force between colliding viscoelastic bodies and finally in Sec.~VII we summarize our findings.

\section{Equation of motion for viscoelastic medium}
When two viscoelastic bodies are brought in a contact, so that
they are deformed,  an interaction force  between the bodies arise.
Generally, it contains  elastic and viscous parts; for a static  contact
however, only the elastic force appears. To compute the forces, one needs to
find a stress that emerges in the  bodies and integrate the stress  over the contact
zone. The distribution of stress in the material is governed by the equation for a continuum medium which reads, e.g.~\cite{Landau:1965},
\begin{equation}
\label{eq:1}
\rho \ddot{\bf u} = {\bf \nabla } \cdot \hat{\sigma}= {\bf \nabla } \cdot  \left( \hat{\sigma}^{ el} + \hat{\sigma}^{ v} \right) \, .
\end{equation}
Here $\rho$ is the material density, ${\bf u}={\bf u}({\bf r})$
is the displacement field in a point ${\bf r}$ and $\hat{\sigma}$
is the stress tensor, comprised of the elastic $\hat{\sigma}^{
el}$ and viscous part $\hat{\sigma}^{ v}$. The elastic stress
linearly depends on the strain tensor, $$u_{ij}=\frac12
\left(\nabla_i u_j+\nabla_j u_i \right), $$ built on the displacement field~\cite{Landau:1965}:
\begin{equation}
 \label{eq:sigma__el}
 \sigma^{el}_{ij} ({\bf u})=2E_1
\left(u_{ij} -\frac13 \delta_{ij} u_{ll}\right) +E_2 \delta_{ij}
u_{ll}\,.
\end{equation}
Similarly, the viscous stress linearly depends on the strain rate
tensor~\cite{Landau:1965}:
\begin{equation}
 \label{eq:sigma_vis}
 \sigma^{v}_{ij} (\dot{\bf u})=2\eta_1
\left(\dot{u}_{ij} -\frac13 \delta_{ij} \dot{u}_{ll}\right) +\eta_2 \delta_{ij}
\dot{u}_{ll}\, .
\end{equation}
Here $E_1= \frac{Y}{2(1+\nu)}$ and  $E_2= \frac{Y}{3(1-2\nu)}$, with
$Y$ and $\nu$ being respectively the Young modulus and Poisson
ratio of the body material. $\eta_1$ and $\eta_2$ are the viscosity coefficients for
the shear and bulk viscosity and  $i,j,l$ denote Cartesian coordinates; the Einstein's summation rule is applied.

The elastic deformation implies that, after separation of the
contacting particles, they completely recover their initial shape so that no plastic deformation remains. Only such deformations will be addressed below.

\begin{figure}[t]
\center{
\includegraphics[width=0.3\textwidth]%
 {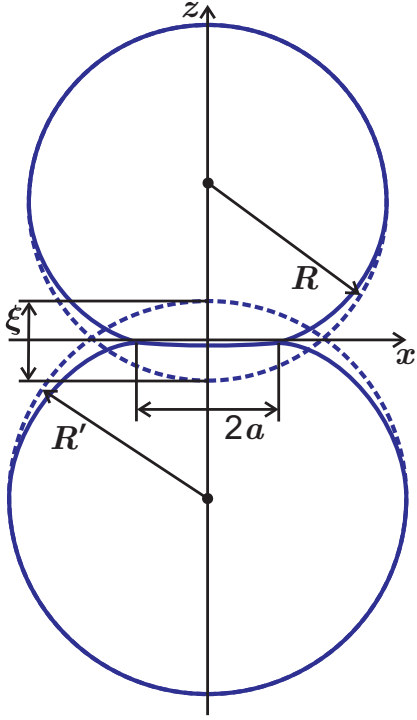}}

  \caption{
Illustrates a simple case of a collision of two visco-elastic spheres (the dashed profiles show undeformed bodies) in the according coordinate frame.  Note however, that in the text a general case of arbitrary convex bodies is addressed.
 }
  \label{fig1}
\end{figure}

\section{Static contact. Hertz theory.}
To introduce the notations and illustrate the main technical ideas,
we start from the simplest case of a static contact, that is, we consider a
time-independent contact of two convex bodies. We assume that only normal
forces, with respect to the contact area, act between the
particles. We place the coordinate system in the center of the
contact region, where $x=y=z=0$ (Fig. \ref{fig1}). Let the displacement field in the
upper body, located at $z>0$, be ${\bf u}({\bf r})$, while in
the lower body, located at $z<0$ be ${\bf w}({\bf r})$. Then the
deformation $\xi$ which is  equal to the sum of the compressions of the
both bodies in the center of the contact zone is related to the
$z$-components of the displacements of the upper and lower bodies' surfaces at the contact
plane $u_{z}(x,y,0)$ and $w_{z}(x,y,0)$,  Fig. \ref{fig1}. It may be shown~\cite{Landau:1965}  that for the
bodies of arbitrary shape the following relation holds true:
\begin{equation}
\label{x2y2uz1uz2}
B_1x^2 +B_2y^2 +u_{z}(x,y,0)+w_{z}(x,y,0) =\xi \,,
\end{equation}
where the constants $B_1$ and $B_2$ are
related to the radii of curvature of the bodies' surfaces near the
contact~\cite{Landau:1965},
\begin{eqnarray}
  \label{B1B2_R1R2}
    2\left( B_1+B_2\right) &= & \frac{1}{R_1}+\frac{1}{R_2}+\frac{1}{R_1^{\prime}} + \frac{1}{R_2^{\prime}}\,, \\
    4\left( B_1\!-\!B_2\right)^2 &= &\left(\frac{1}{R_1}-\frac{1}{R_2}\right)^2
    +\left(\frac{1}{R_1^{\prime}}-\frac{1}{R_2^{\prime}}\right)^2
    \nonumber\\
    &&\hspace{-7mm} {} + 2 \cos 2 \varphi \left(\frac{1}{R_1} - \frac{1}{R_2}\right)
    \left(\frac{1}{R_1^{\prime}} - \frac{1}{R_2^{\prime}} \right).
\end{eqnarray}
Here $R_1$, $R_2$ and $R_1^{\prime}$, $R_2^{\prime}$ are
respectively the principal radii of curvature of the first and the
second body at the point of contact and  $\varphi$ is the angle
between the planes corresponding to the curvature radii $R_1$ and
$R_1^{\prime}$. Equations~(\ref{x2y2uz1uz2}) and (\ref{B1B2_R1R2})
describe the general case of the contact between two smooth bodies
(see \cite{Landau:1965} for details). The physical meaning of
\eqref{x2y2uz1uz2} is easy to see for the case of a contact of a
soft sphere of a radius $R$ ($R_1=R_2=R$) with a hard, undeformed
plane ($R_1^{\prime}=R_2^{\prime}=\infty$). In this case
$B_1=B_2=1/2R$, the compressions of the sphere and of the plane are
respectively $u_{z}(0,0,0)=\xi$ and $w_{z}=0$, and the surface of
the sphere before the deformation is given by
$z(x,y)=(x^2+y^2)/2R$ for small $z$. Then \eqref{x2y2uz1uz2} reads in the
flattened area $u_{z}(x,y)= \xi - z(x,y)$, that is, it gives  the
condition for a point $z(x,y)$ on the body's surface to touch the
plane $z=0$.
While Eq.~(\ref{x2y2uz1uz2}) defines the displacement on the
contact surface, the displacement fields in the bulk of the first
(upper) and second (lower) bodies are determined by the following
equations.
\begin{equation}
\label{eq:2c}
{\bf \nabla} \cdot \hat{\sigma}^{el} ({\bf u})=0,
\qquad \qquad
{\bf \nabla} \cdot \hat{\sigma}^{el} ({\bf w})=0.
\end{equation}

Both equations may be solved by the same approach, therefore  in what follows we
consider the solution for the upper body with
$z>0$. Using Eq.~(\ref{eq:sigma__el}) which relates the stress and
strain tensors, we write:
\begin{equation}
\nabla_j\sigma_{ij}^{el}
  =E_1\Delta u_i+\left(E_2+\frac{1}{3}E_1\right)\nabla_i \nabla_j u_j=0\,,
\label{eq04}
\end{equation}
where the elastic constants refer to the upper body (for the
notation simplicity we do not add now the additional index specifying the body -- it will be done later).

To solve the above equation we use the approach
of~\cite{Landau:1965} and write the  solution as
\begin{equation}
{\bf u}^{(0)}=f^{(0)}{\bf e}_z+\nabla\varphi^{(0)}\,,
\label{eq05}
\end{equation}
where $\varphi^{(0)}=K^{(0)}zf^{(0)}+\psi^{(0)}$, $K^{(0)}$ is
some constant to be found and  $f^{(0)}$ and  $\psi^{(0)}$ are
unknown harmonic functions. We assume the lack of tangential
stress at the interface, which is e.g.\ fulfilled when the bodies
at a contact are of the same material. Taking into account that
\begin{equation}
\displaystyle\Delta {\bf u}=\Delta\nabla\varphi^{(0)}
 =2K^{(0)}\nabla\frac{\partial f^{(0)}}{\partial z}
\label{eq:5a}
\end{equation}
and
\begin{equation}
 \displaystyle\nabla\cdot {\bf u}=(1+2K^{(0)})\frac{\partial f^{(0)}}{\partial z},
\label{eq:5b}
\end{equation}
we recast Eq.~(\ref{eq04}) into the following form:
\begin{eqnarray}
\nabla_j\sigma_{ij}^{el} &=& \bigg[2E_1 K^{(0)}+\\
 &+ & (1+2K^{(0)})\left(E_2+\frac{E_1}{3}\right)\bigg]
 \nabla_i\frac{\partial f^{(0)}}{\partial z}=0, \nonumber
\end{eqnarray}
which implies that
\begin{equation}
K^{(0)}=-\frac{1}{2}\frac{3E_2+E_1}{3E_2+4E_1}\,.
\label{eq07}
\end{equation}
Consider now the boundary condition for the stress tensor.
Obviously, on the free boundary all components of the stress
vanish. In the contact region, located at the surface, $z=0$, the
tangential components of the stress tensor $\sigma_{zx}^{el}$ and
$\sigma_{zy}^{el}$ vanish as well, while the normal component of
the stress tensor reads,
\begin{equation}
{\bf n} \cdot \hat{\sigma}^{el} =-\sigma_{zz}^{el}=P_z
\label{eq:sigmaPz}
\end{equation}
where ${\bf n}=(0,0,-1)$ is the external normal to the upper body
on the contact plane and $P_z$ is the normal pressure acting on
the contact surface. Therefore the boundary conditions have the
following form:
\begin{equation}
\left.\sigma_{zx}^{el}\right|_{z=0}\!=\!0; \quad \left.\sigma_{zy}^{el}\right|_{z=0}\!=\!0; \quad \left.\sigma_{zz}^{el}\right|_{z=0}\!=\!-P_z.
\label{eq10}
\end{equation}
Using the expression (\ref{eq:sigma__el}) for the elastic part of
the stress tensor, together with the displacement vector
(\ref{eq05}) we recast the boundary conditions (\ref{eq10}) into
the form:
\begin{eqnarray}
\label{eq12}
&&\left.\frac{\partial}{\partial x}\!\left(\frac{3E_1 }{4E_1+3E_2}f^{(0)}
 \!+\!2\frac{\partial\psi}{\partial z}\right)\right|_{z=0} \!\!=0 \\
\label{eq13}
&&\left.\frac{\partial}{\partial y}\!\left(\frac{3E_1 }{4E_1+3E_2}f^{(0)}
 \!+\!2\frac{\partial\psi}{\partial z}\right)\right|_{z=0} \!\!=0 \\
\label{eq14}
&&\left.\frac{\partial}{\partial z}\!\left(\frac{3E_1 }{4E_1 +3E_2}f^{(0)}
 \!+\!2\frac{\partial\psi}{\partial z}\right)\right|_{z=0}\!\!
 \!=\!- \frac{P_z}{E_1}.
\end{eqnarray}
From equations (\ref{eq12}) and (\ref{eq13}) follows the relation
between $f^{(0)}$ and $\frac{\partial\psi}{\partial z}$ at $z=0$:
\begin{equation}
\label{eq14a}
\left.\left(\frac{\partial\psi}{\partial z}
 +\frac{3}{2}\frac{E_1}{4E_1+3E_2}f^{(0)}\right)\right|_{z=0}
 ={\rm const} =0\,.
\end{equation}
The constant in the above relation equals to zero, since it holds
true independently on the coordinate that is, also at the
infinity; at the infinity, however, the deformation and thus the
above functions vanish. Since $f^{(0)}$, $\psi$ as well as
$\partial\psi/\partial z$ are the harmonic functions, the
condition that their  linear combination  vanishes on the
boundary, Eq.~(\ref{eq14a}), implies that it is zero in the total
domain, that is,
\begin{equation}
\frac{\partial\psi}{\partial z}
 =-\frac{3}{2}\frac{E_1}{4E_1+3E_2}f^{(0)}\,.
\label{eq15}
\end{equation}
Substituting the last relation into (\ref{eq14}) yields
\begin{equation}
\left.\frac{\partial f^{(0)}}{\partial z}\right|_{z=0}
 =-\frac{4E_1+3E_2}{E_1(E_1 +3E_2)}P_z   .
\label{eq16}
\end{equation}
Since $f^{(0)}$ is a harmonic function, one can use the relation
between the normal derivative of a harmonic function on the
surface and its value in the bulk, as it follows from the theory
of harmonic functions (see e.g.~\cite{Landau:1965,Tikhonov}),
hence we find:
\begin{equation}
f^{(0)}({\bf r})=\frac{4E_1+3E_2}{2\pi E_1(E_1+3E_2)}
 \int\!\!\!\int_S\frac{ P_z(x',y')\,\mathrm{d}x'\mathrm{d}y'}{ \left|{\bf r}
 -{\bf r}'\right|}\,,
\label{eq17}
\end{equation}
where $S$ is the contact area. Using Eq.~(\ref{eq05}) we can write
$z$-component of the zero-order displacement at $z=0$ as
$$
\displaystyle\left.u_z\right|_{z=0}=(1+K^{(0)})\left.f^{(0)}\right|_{z=0}
 +\left.\frac{\partial\psi}{\partial z}\right|_{z=0},
$$
which together with (\ref{eq15}) and definition of $K^{(0)}$,
Eq.~(\ref{eq07}) yields,
\begin{equation}
\left.u_z\right|_{z=0}=\frac{1}{2}\left.f^{(0)}\right|_{z=0} \,.
\label{eq20}
\end{equation}
If we now express $E_1$ and $E_2$ in terms of $\nu_1$ and $Y_1$,
where $\nu_1$ and $Y_1$ are the according constants for the upper
body (recall that we consider the upper contacting body) one
obtains from Eqs.~(\ref{eq20}), (\ref{eq17}) and (\ref{eq10}):
\begin{equation}
\label{eq20a}
u_z(x,y, z=0)=-\frac{(1-\nu_1^2)}{\pi Y_1} \int\!\!\!\int_S
\frac{ \sigma_{zz}^{el}(x',y', z=0)\,\mathrm{d}x'\mathrm{d}y'}{ \left|{\bf r} -{\bf r}'\right|}\,.
\end{equation}
The same considerations may be performed for the lower body.
Taking into account that the external normals for the upper and
lower bodies as well as the exerted pressures are equal up to a
minus sign
 (${\bf n}_{\rm up} =- {\bf n}_{ \rm l}$, $P_z=P_{z, \rm up} =-P_{z, \rm l} $),
we obtain,
\begin{equation}
\label{eq20ab}
w_z(x,y, z=0) =-\frac{(1-\nu_2^2)}{\pi Y_2} \int\!\! \int_S
\frac{ \sigma_{zz}^{el}(x',y', z=0)\,\mathrm{d}x'\mathrm{d}y'}{ \left|{\bf r} -{\bf r}'\right|}\,.
\end{equation}
Hence, with Eq.~(\ref{eq:sigmaPz}) the relation (\ref{x2y2uz1uz2})
takes the form:
\begin{eqnarray}
\frac{1}{\pi} \left( \frac{1-\nu_1^2}{Y_1} +  \frac{1-\nu_2^2}{Y_2} \right)
 \int\!\!\!\int_S \frac{P_z(x^{\prime},y^{\prime})}{\left|{\bf r} -{\bf r}'\right|}dx^{\prime}d y^{\prime}=
 \quad &&\nonumber\\[5pt]
 = \xi -B_1x^2-B_2y^2 \,,&&
 \label{Eq_for_uz12}
\end{eqnarray}
The equation \eqref{Eq_for_uz12} is an integral equation for the
unknown function $P_z(x,y)$. We compare this equation with the
mathematical identity~\cite{Landau:1965}
\begin{eqnarray}
&&\hspace{-7mm}
\int\!\!\!\int_S  \frac{\mathrm{d}x^{\prime}\mathrm{d}y^{\prime}}{|{\bf r} - {\bf r}^{\prime}|} \sqrt{1\!-\!\frac{x^{\prime\, 2}}{a^2} \!-\!\frac{y^{\prime \, 2}}{b^2}}=
\\
 && \qquad \hspace{-7mm}=
\frac{\pi ab}{2} \!\int\limits_0^{\infty} \left[\! 1\!-\!\frac{x^2}{a^2+t}\!-\!\frac{y^2}{b^2+t} \right]
\!\frac{dt}{\sqrt{(a^2+t)(b^2+t)t}}\,,\nonumber
\label{identity}
\end{eqnarray}
where integration is performed over the elliptical area \\
$x^{\prime\, 2}/a^2+y^{\prime\, 2}/b^2\le1$. The left-hand sides
of the both equations, \eqref{Eq_for_uz12} and  \eqref{identity}, contain integrals of the same type, while the right-hand sides contain quadratic forms of the same type.
Therefore, the contact area is an ellipse with the semi-axes $a$
and $b$ and the pressure is of the form $$P_z(x,y) ={\rm const}
\sqrt{1-x^2/a^2-y^2/b^2}.$$ The constant here may be found from the
total elastic force $F_\el$ acting between the bodies. Integrating
$P_z(x,y)$ over the contact area we get $F_\el$, which then yields the constant. Hence we obtain
\begin{equation}
\label{Pressure}
P_z(x,y) =\frac{3F_\el}{2\pi ab} \sqrt{1-\frac{x^2}{a^2} -\frac{y^2}{b^2}} \,.
\end{equation}
We substitute \eqref{Pressure} into \eqref{Eq_for_uz12} and
replace the double integration over the contact area by
integration over the variable $t$, according to the above identity.   Thus, we obtain an equation containing terms
proportional to $x^2$, $y^2$ and a constant. Equating the
corresponding coefficients we obtain
\begin{eqnarray}
\label{xi_Fel}
&&\hspace{-5mm}
\xi=\frac{F_\el D}{\pi} \int_0^{\infty} \frac{\mathrm{d}t}{\sqrt{(a^2+t)(b^2+t)t}} =\frac{F_\el D}{\pi}\frac{N(\zeta)}{b}\,,\\[5pt]
\label{B1_Fel}
&&\hspace{-5mm}
 B_1=\frac{F_\el D}{\pi} \int_0^{\infty} \frac{\mathrm{d}t}{(a^2+t) \sqrt{(a^2+t)(b^2+t)t}}
\nonumber\\
&&\qquad=\frac{F_\el D}{\pi} \frac{M(\zeta)}{a^2b}\,,\\[5pt]
\label{B2_Fel}
&&\hspace{-5mm}
 B_2=\frac{F_\el D}{\pi} \int_0^{\infty} \frac{\mathrm{d}t}{(b^2+t) \sqrt{(a^2+t)(b^2+t)t}} \,
\nonumber\\
&&\qquad=\frac{F_\el D}{\pi} \frac{M(1/\zeta)}{ab^2}\,,
\label{B3x}
\end{eqnarray}
where
\begin{equation}
\label{def_D}
D \equiv \frac34 \left( \frac{1-\nu_1^2}{Y_1} + \frac{1-\nu_2^2}{Y_2} \right)
\end{equation}
and $\zeta \equiv a^2/b^2$ is the ratio of the contact ellipse
semi-axes. In \eqref{xi_Fel}--\eqref{B3x} we introduce the
short-hand notations\footnote{The function $N(\zeta)$ and
$M(\zeta)$ may be expressed as a combination of the Jacobian
elliptic functions $E(\zeta)$ and $K(\zeta)$
\cite{AbramowitzStegun:1965}.}
\begin{eqnarray}
\label{eq:def_N}
 &&N(\zeta)=\int_0^{\infty} \frac{\mathrm{d}t}{\sqrt{(1+\zeta t)(1+t)t}} \\
\label{eq:def_M}
 &&M(\zeta)=\int_0^{\infty} \frac{\mathrm{d}t}{(1+t)\sqrt{(1+t)(1+\zeta t)t}}\, .
\end{eqnarray}
From the above relations follow the size of the contact area,
$a$, $b$  and the deformation $\xi$ as functions of the elastic
force $F_\el$ and (known) geometric coefficients $B_1$ and $B_2$.

The dependence of the force $F_\el$ on the deformation $\xi$ may
be obtained from scaling arguments. If we rescale
 $a^2 \to \alpha a^2$, $b^2 \to \alpha b^2$, $\xi \to \alpha \xi$ and $F_\el \to \alpha^{3/2} F_\el$,
with $\alpha$ constant, Eqs.~(\ref{xi_Fel})--(\ref{B2_Fel}) remain
unchanged. That is, when $\xi$ changes by the factor $\alpha$, the
semi-axis $a$ and $b$ change by the  factor $\alpha^{1/2}$ and the
force by the factor
 $\alpha^{3/2}$, i.e., $a \sim \xi^{1/2}$, $b \sim \xi^{1/2}$ and
\begin{equation}
\label{Fel_xi_gen}
F_\el = {\rm const } \, \xi^{3/2} \, .
\end{equation}
The  dependence \eqref{Fel_xi_gen} holds true for all smooth
convex bodies in contact. To find the
constant in \eqref{Fel_xi_gen} we divide \eqref{B2_Fel} by
\eqref{B1_Fel} and obtain the transcendental equation
\begin{equation}
\label{eq:k}
\frac{B_2}{B_1}= \frac{\sqrt{\zeta} M \left(1/\zeta \right)}{M(\zeta)}
\end{equation}
for the ratio of semi-axes $\zeta$. Let $\zeta_0$ be the root of
Eq.~\eqref{eq:k}, then $a^2=\zeta_0b^2$ and we obtain from Eqs.~\eqref{xi_Fel}, \eqref{B1_Fel}:
\begin{eqnarray}
\label{eq:xi_N0}
&&\xi=\frac{F_\el D}{\pi}\frac{N(\zeta_0)}{b} \\
\label{eq:B1_M0}
&&B_1=\frac{F_\el D}{\pi}\frac{M(\zeta_0)}{\zeta_0b^3} \, ,
\end{eqnarray}
where $N(\zeta_0)$ and $M(\zeta_0)$ are pure numbers.
Equations~(\ref{eq:xi_N0}), (\ref{eq:B1_M0}) allow us to find the
semi-axes $b$ and the elastic force $F_\el$ as functions of the
compression $\xi$. Hence we obtain the force, that is, we get the
according constant in Eq.~\eqref{Fel_xi_gen} ~\cite{BP_Willey2004}:
\index{elastic interaction force}
\begin{equation}
\label{Fel_xi_gen1}
 F_\el = \frac{\pi}{D} \left( \frac{M(\zeta_0)}{B_1 \zeta_0 N(\zeta_0) } \right)^{1/2} \, \xi^{3/2} = C_0 \xi^{3/2}\,.
\end{equation}
Similarly we can relate the deformation $\xi$ and the semi-axes
$a$ of the contact ellipse~\cite{BP_Willey2004}:
\begin{equation}
\label{eq:xi_a}
 a= \left( \frac{M(\zeta_0)}{N(\zeta_0)B_1} \right)^{1/2} \xi^{1/2}\,.
\end{equation}
Note that $\zeta_0$ is a constant determined by the collision
geometry.

For the special case of contacting spheres of the same material ($a=b$), the constants $B_1$ and $B_2$ read
\begin{equation}
\label{Spheres}
B_1=B_2=\frac12 \left( \frac{1}{R_1}+\frac{1}{R_2} \right)=\frac12 \frac{1}{R^{\rm eff}} \, .
\end{equation}
In this case $\zeta_0=1$, $N(1)=\pi$, and $M(1)=\pi/2$, leading to
the solution of (\ref{eq:xi_N0}), (\ref{eq:B1_M0}):
\index{Hertz contact law}
\begin{eqnarray}
\label{eq:a2_Hertz}
&&a^2=R^{\rm eff} \, \xi \\
\label{Fel_xi_sph}
&&F_\el = \rho \xi^{3/2} \, ;  \qquad \rho \equiv \frac{2Y}{3(1-\nu^2)} \sqrt{R^{\rm eff}} \, ,
\end{eqnarray}
where we use the definition \eqref{def_D} of the constant $D$.
This contact problem was solved by Heinrich Hertz in 1882
\cite{Hertz:1882}. It describes the force between {\em elastic}
particles. For inelastically deforming particles it describes the
repulsive force in the static case.

\section{Dynamical contact. Perturbation scheme}
\index{viscous interaction force}
For the most important applications the viscous forces are
significantly smaller than the elastic forces and the bodies
material is rigid enough to neglect inertial effects for
collisions with not very large velocities. Let us estimate the
magnitude of different terms in Eq.~(\ref{eq:1}). This may be
easily done using the dimensionless units. For the length scale we
take $R$, which corresponds to the characteristic size of
colliding bodies, while for the time scale we use $\tau_c$ -- the
collision duration.  Then $v_0=R/\tau_c$ is the characteristic
velocity at the impact. Taking into account that differentiation
with respect to a coordinate yields for dimensionless quantities
the factor $1/R$, and with respect to time -- $1/\tau_c$, we
obtain
\begin{eqnarray}
\label{eq:2a}&&
\nabla \sigma^{v} \sim \lambda_1 \,\nabla \sigma^{el} \qquad \qquad \lambda_1 = \tau_{rel}/ \tau_c\,, \\
\label{eq:2b}&&
\rho \ddot{ u} \sim \rho \ddot{ w} \sim \lambda_2 \, \nabla \sigma^{el} \qquad \; \lambda_2 = v_0^2/c^2\,.
\end{eqnarray}
Here $c^2=Y/\rho$ and $\tau_{rel}=\eta/Y$ characterize
respectively the speed of sound and the microscopic relaxation
time in the material and
 $\eta \sim \eta_{1}\sim \eta_{2}$~\cite{bshp96}.

Neglecting terms, of the order of $\lambda_1$ and $\lambda_2$ we get
\begin{equation}
\label{eq:2c}
{\bf \nabla} \cdot \hat{\sigma}^{el} ({\bf u})=0,
\qquad \qquad {\bf \nabla} \cdot \hat{\sigma}^{el} ({\bf w})=0,
\end{equation}
which yields the static displacement fields ${\bf u} ={\bf u}({\bf
r})$ and ${\bf w} ={\bf w}({\bf r})$. This approximation
corresponds to the quasi-static approximation, used in the
literature~\cite{bshp96,Zheng1,Zheng2,bri2,Dintwa2004}.
Neglecting terms of the order $\lambda_2$ but keeping these of the
order of $\lambda_1$, leads to the following equation
\begin{equation}
\label{eq:3}
 {\bf \nabla}  \cdot \left( \hat{\sigma}^{el} ({\bf u})+ \hat{\sigma}^{v} (\dot{{\bf u}}) \right)=0,
\quad
 {\bf \nabla}  \cdot \left( \hat{\sigma}^{el} ({\bf w})+ \hat{\sigma}^{v} (\dot{{\bf w}}) \right)=0.
\end{equation}
That is, to go beyond the quasi-static approximation one needs to
find the solution of Eq.~(\ref{eq:3}) which contains both the
displacement fields ${\bf u}$,  ${\bf w}$, as well as its time
derivatives, $\dot{{\bf u}}$, $\dot{{\bf w}}$. Eq.~(\ref{eq:3})
needs to be supplemented by the boundary conditions. These
correspond to vanishing stress on the free surface and given
displacement in the contact area.

In a vast majority of applications $\lambda_1= \tau_{rel}/\tau_c
\ll 1$, which implies that the viscous stress is small as compared
to the elastic stress. This allows to solve Eq.~(\ref{eq:3})
perturbatively, as a series in a small parameter. Here we follow
the standard perturbation scheme,
e.g.~\cite{book}: To notify the order of
different terms  we introduce a "technical" small parameter
$\lambda$, which at the end of computations is to be taken as one.
Hence one can write,
\begin{equation}
\label{eq:4}
\hat{\sigma} = \hat{\sigma}^{(0)} + \lambda \hat{\sigma}^{(1)} + \lambda^2 \hat{\sigma}^{(2)} + \ldots
\end{equation}
and respectively,
\begin{eqnarray}
\label{eq:4a}
&&{\bf u}({\bf r})= {\bf u}^{(0)}({\bf r})+\lambda {\bf u}^{(1)}({\bf r})+ \lambda^2 {\bf u}^{(2)}({\bf r}) +\ldots, \\
\label{eq:4aa}
&&{\bf w}({\bf r})= {\bf w}^{(0)}({\bf r})+\lambda {\bf w}^{(1)}({\bf r})+ \lambda^2 {\bf w}^{(2)}({\bf r}) +\ldots
\end{eqnarray}

Substituting the expansions (\ref{eq:4}) and (\ref{eq:4a}),
(\ref{eq:4aa}) into Eq.~(\ref{eq:3}) yields a set of equations for
different order in $\lambda$. Zero-order equations with the
according boundary conditions  read,
\begin{eqnarray}
\label{eq:5}
 &&{\bf \nabla}  \cdot  \hat{\sigma}^{el} \left({\bf u}^{(0)}\right)=0\,,
\quad \qquad {\bf \nabla}  \cdot  \hat{\sigma}^{el} \left({\bf w}^{(0)}\right)=0\,,
\\
&&B_1x^2 +B_2y^2 +u_{z}^{(0)}(x,y,0)+w_{z}^{(0)}(x,y,0)=\xi \,,
\nonumber
\end{eqnarray}
while the first-order equations with the boundary conditions have the form
\begin{eqnarray}
\label{eq:6}
 &&{\bf \nabla}  \cdot \left( \hat{\sigma}^{el} ({\bf u}^{(1)})+ \hat{\sigma}^{v} (\dot{{\bf u}}^{(0)}) \right)=0\,, \nonumber\\
&&  {\bf \nabla}  \cdot \left( \hat{\sigma}^{el} ({\bf w}^{(1)})+ \hat{\sigma}^{v} (\dot{{\bf w}}^{(0)}) \right)=0\,,
 \\
&&  u_z^{(1)}(x,y,0)+w_z^{(1)}(x,y,0)=0 \,,
\nonumber
\end{eqnarray}
and so on.  Note that the zero-order equation (\ref{eq:5})
corresponds to case of a static contact which has been considered
in detail above. This also corresponds to the  quasi-static
approximation widely used in the literature,
e.g.~\cite{bshp96,Zheng1,Zheng2,bri2,Dintwa2004}.
Also note that in the proposed perturbation scheme, only
zero-order problem (\ref{eq:5}) has non-zero boundary conditions,
corresponding to the boundary conditions (\ref{x2y2uz1uz2}) of the
initial problem; all other, high-order  perturbation equations,
have homogeneous boundary conditions. Such partition of the
boundary conditions is justified due to the linearity of the
problem.

Note that for the zero-order solution the condition
$\sigma_{zz}^{el}({\bf u}^{(0)}) = \sigma^{el}({\bf w}^{(0)})$ is
fulfilled at the contact plane $z=0$, as it directly follows from the construction of
the solution. For the first-order solution, however, we need to
additionally request the  condition for the first-order
stress tensor:
\begin{eqnarray}
&&\hspace{-10mm}
\left. \left( \sigma_{zz}^{v }({\bf u}^{(0)})  +
 \sigma_{zz}^{el}({\bf u}^{(1)}) \right)\right|_{z=0}
 \nonumber\\
\label{eq:consists}
&&\qquad =
\left. \left( \sigma_{zz}^{v }({\bf w}^{(0)}) +
 \sigma_{zz}^{el}({\bf w}^{(1)}) \right) \right|_{z=0}\,,
\end{eqnarray}
which implies the equivalence of the first-order stress tensor,
expressed in terms of deformation and deformation rate of the
upper body and of the lower one.

\section{First-order solution. Beyond quasi-static approximation.}
Again we will consider the upper body with $z>0$ and introduce,
for convenience, the following notations:
\begin{eqnarray}
&&
\hat{\sigma}^{el} \left({\bf u}^{(0)}\right) = \hat{\sigma}^{el\,(0)},
\qquad
\hat{\sigma}^{el} \left({\bf u}^{(1)}\right) = \hat{\sigma}^{el\,(1)},
\nonumber
\\
&&
\hat{\sigma}^{v} \left(\dot{\bf u}^{(0)}\right) = \hat{\sigma}^{v\,(1)},
\qquad \mbox{etc.}
\nonumber
\end{eqnarray}
With this notations and using Eqs.~(\ref{eq:sigma__el}),
(\ref{eq:sigma_vis}) and (\ref{eq:5b}) we write:
\begin{eqnarray}
 \sigma_{ij}^{v} \left(\dot{\bf u}^{(0)}\right)&=&\sigma_{ij}^{v(1)} \\
&=&\frac{\eta_1}{E_1}\dot{\sigma}_{ij}^{el(0)}+\left(\eta_2\!-\!\eta_1\frac{E_2}{E_1}\right)(1\!+\!2K^{(0)})
 \frac{\partial\dot{f}^{(0)}}{\partial z}\delta_{ij}\,, \nonumber
\label{eq09}
\end{eqnarray}
and accordingly the divergence of this tensor:
\begin{eqnarray}
\nabla _j\sigma_{ij}^{v(1)}&=&\left[2\eta _1 K^{(0)}
 +(1+2K^{(0)})\left(\eta _2+\frac{\eta _1}{3}\right)\right]
 \nabla_i\frac{\partial\dot{f}^{(0)}}{\partial z} \nonumber \\
 &=& \frac{3(E_1\eta_2 -E_2 \eta_1)}{(4 E_1+3E_2)} \nabla_i\frac{\partial\dot{f}^{(0)}}{\partial z}\,,
\label{eq08}
\end{eqnarray}
where Eqs.~(\ref{eq:5a}), (\ref{eq:5b}) and Eq.~(\ref{eq07}) for
$K^{(0)}$ have been used. If we now apply Eq.~(\ref{eq16}) for
$\partial\dot{f}^{(0)}/\partial z$ and again Eq.~(\ref{eq07}) for
the constant $K^{(0)}$, we find the $zz$-component of the
first-order dissipative tensor on the contact plane, $z=0$:
\begin{eqnarray}
\label{eq19}
\sigma_{zz}^{v(1)}(x,y,0)&=& \alpha  \dot{\sigma}_{zz}^{el(0)}(x,y,0) \\
\label{eq19a}
\alpha &=& \frac{3\eta_2+\eta_1}{E_1+3E_2}.
\end{eqnarray}
Similar relation may be obtained for the lower body. Using the
definitions of $E_1$ and $E_2$ the coefficient $\alpha$ reads
for each of the bodies,
\begin{equation}
\label{eq:alpha0}
\alpha_i =\frac{(1+\nu_i)(1-2\nu_i)}{Y_i}\left( 2\eta_{2(i)} + \frac23 \eta_{1 (i)} \right),
\end{equation}
where the subscript $i=1,2$ specifies the body -- $i=1$ for
the upper body and $i=2$ for the lower one. The above relation
corresponds to the according approximation of
Ref.~\cite{Zheng1,Zheng2} and coincides
with the result of~\cite{Zheng1,Zheng2}, where the necessary
corrections have been introduced. Note, however, that quasi-static
approximation occurs to be inconsistent for the case of contact of
particles of different material: Indeed, the condition (\ref{eq:consists}) is possible only if the first-order elastic terms are taken into account. Obviously, this may not be achieved within the quasi-static approximation, which uses only the first-order dissipative stress $\sigma_{zz}^{v(1)}$. The values of $\sigma_{zz}^{v(1)}$ on the contact plane are different for the upper and lower body for different materials, since $\alpha_1 \neq \alpha_2$ [see Eqs.~(\ref{eq19})-(\ref{eq:alpha0})],  that is, the third Newton's law for this case is violated.

Consider now  the first-order equation (\ref{eq:6}):
\begin{equation}
\nabla_j(\sigma_{ij}^{el(1)}+\sigma_{ij}^{v(1)})=0\,.
\label{eq21}
\end{equation}
Due to the linearity of the problem, one can represent the
first-order displacement field as a sum of two parts, ${\bf
u}^{(1)} = \bar{\bf u}^{(1)}+\tilde{\bf u}^{(1)}$, which
correspond to the two parts of the elastic tensor,
$\sigma_{ij}^{el(1)}=\tilde{\sigma}_{ij}^{el(1)}(\tilde{\bf
u}^{(1)})+\bar{\sigma}_{ij}^{el(1)}(\bar{\bf u}^{(1)})$. Here the
first part of $\sigma_{ij}^{el(1)}$ is the solution of the
\emph{inhomogeneous} equation with homogeneous boundary
conditions:
\begin{eqnarray}
\label{eq22}
&&\nabla_j\tilde{\sigma}_{ij}^{el(1)}=-\nabla_j\sigma_{ij}^{v(1)}\,,  \\
\label{eq22a}
&&\left.\tilde{\sigma}_{xz}^{el(1)}\right|_{z=0}=
 \left.\tilde{\sigma}_{yz}^{el(1)}\right|_{z=0}=
 \left.\tilde{\sigma}_{zz}^{el(1)}\right|_{z=0}=0\,,
\end{eqnarray}
while the second part is the solution of the \emph{homogeneous} equation,
\begin{equation}
\label{eq23}
\nabla_j\bar{\sigma}_{ij}^{el(1)}=0\,,
\end{equation}
with a given first-order displacement ${u}_z^{(1)}$ at the contact plane; this is to be obtained from the boundary condition (\ref{eq:6}) and consistency condition (\ref{eq:consists}). The boundary problem (\ref{eq23}) is exactly the same as the above problem (\ref{eq:5}) for the zero-order functions. Hence the same relation (\ref{eq20a}) holds true for the first-order functions, that is,
\begin{eqnarray}
&&\hspace{-10mm}
\left.\bar{u}_z^{(1)}\right|_{z=0}=-\frac{(1-\nu_1^2)}{\pi Y_1} \\
\label{eq:29b}
&&~~\times \int\!\!\!\int_S
\frac{ \bar{\sigma}_{zz}^{el(1)}(\bar{\bf u}^{(1)}(x',y', z=0))\,\mathrm{d}x'\mathrm{d}y'}{ \left|{\bf r} -{\bf r}'\right|}\,. \nonumber
\end{eqnarray}

To solve Eq.~(\ref{eq22}) we write the displacement field
${\tilde{\bf u}}^{(1)}$ in a form, similar to this of the
zero-order solution (\ref{eq05}):
\begin{equation}
{\bf \tilde{u}}^{(1)}=f^{(1)}{\bf e}_z+\nabla\varphi^{(1)}\,,
\label{eq24}
\end{equation}
where $\varphi^{(1)}=K^{(1)}zf^{(1)}+\psi^{(1)}$, $K^{(1)}$ is
some constant and  $f^{(1)}$ and  $\psi^{(1)}$ are harmonic
functions. Then we can write the stress tensor
$\tilde{\sigma}_{ij}^{el(1)}$ as
\begin{eqnarray}
 \tilde{\sigma}_{ij}^{el(1)}&=&(1+2K^{(1)})
 \left[E_1(\delta_{jz}\nabla_i f^{(1)}+\delta_{iz}\nabla_jf^{(1)}) \right. + \nonumber \\
 &+& \left.\left(E_2-\frac{2}{3}E_1\right)\frac{\partial f^{(1)}}{\partial z}\delta_{ij}\right]
 {}+2E_1K^{(1)}z\nabla_i\nabla_j f^{(1)} \nonumber \\ &+& 2E_1\nabla_i\nabla_j\psi^{(1)}.
\label{eq25}
\end{eqnarray}
If we choose $K^{(1)}=-\frac12$ the above stress tensor takes the form
\begin{equation}
\label{eq:25a}
\tilde{\sigma}_{ij}^{el(1)}= - z E_1\nabla_i\nabla_j f^{(1)} + 2 E_1 \nabla_i\nabla_j\psi^{(1)}
\end{equation}
and the boundary conditions (\ref{eq22a}) read:
\begin{eqnarray}
\label{eq:25b}
&&\left.\tilde{\sigma}_{xz}^{el(1)}\right|_{z=0}= \frac{\partial}{\partial x} \left.\left( \frac{\partial \psi^{(1)}}{\partial z} \right) \right|_{z=0}=0 \,,\\
&&\left.\tilde{\sigma}_{yz}^{el(1)}\right|_{z=0}= \frac{\partial}{\partial y} \left.\left( \frac{\partial \psi^{(1)}}{\partial z} \right) \right|_{z=0}=0 \,.
\end{eqnarray}
Therefore we conclude,
\begin{equation}
\label{eq:25c}
\left.\frac{\partial \psi^{(1)}}{\partial z}  \right|_{z=0}={\rm const}=0 \,,
\end{equation}
where the last equation follows from the condition that
$\psi^{(1)}$ vanishes at the infinity, $x,\,y \to \infty$, where
the deformation is zero. Since $\psi^{(1)}$ is a harmonic
function, we conclude that the vanishing normal derivative on a
boundary, Eq.~(\ref{eq:25c}), implies that the function vanishes
everywhere, that is, $\psi^{(1)}(x,y,z)=0$ (see
e.g.~\cite{Tikhonov}). Hence
\begin{equation}
\label{eq:25d}
\tilde{\sigma}_{ij}^{el(1)}=-E_1z\nabla_i\nabla_j f^{(1)}
\end{equation}
and the third boundary condition, $\tilde{\sigma}_{zz}^{el(1)}=0$
at $z=0$ is automatically fulfilled. Taking into account that
function $f^{(1)}$ is harmonic, we obtain,
$$
\nabla_j \tilde{\sigma}_{ij}^{el(1)}= -E_1  \nabla_i \frac{\partial f^{(1)} }{\partial z}\,.
$$
Using the above equation together with Eq.~(\ref{eq08}) we recast
Eq.~(\ref{eq22}) into the form,
$$
E_1  \nabla_i \frac{\partial f^{(1)} }{\partial z} =
-\frac{3(E_2 \eta_1 - E_1\eta_2)}{(4 E_1+3E_2)} \nabla_i\frac{\partial\dot{f}^{(0)}}{\partial z}\,
$$
which implies the relation between functions $f^{(1)}$ and
$\dot{f}^{(0)}$:
\begin{eqnarray}
\label{eq28}
f^{(1)} &=&- \beta  \dot{f}^{(0)}\\
\label{eq28al}
\beta &=&\frac{3(E_2\eta_1-E_1\eta_2)}{E_1(3E_2+4E_1)}\,.
\end{eqnarray}
Using Eq.~(\ref{eq24}) with $K^{(1)}=-\frac12$ we write for
$\tilde{u}_z^{(1)}$:
\begin{equation}
\tilde{u}_z^{(1)} =\frac12 f^{(1)} - \frac{z}{2} \frac{\partial f^{(1)} }{\partial z}\,;
\label{eq28a}
\end{equation}
substituting there $f^{(1)}$ from Eq.~(\ref{eq28}) we arrive at
\begin{equation}
\tilde{u}_z^{(1)} = -\frac12 \beta \left( \dot{f}^{(0)} -z \frac{\partial \dot{f}^{(0)} }{\partial z} \right)\,,
\label{eq28b}
\end{equation}
where $f^{(0)}$ is given by Eq.~(\ref{eq17}). Thus, the above
relation presents the solution for  the displacement
$\tilde{u}_z^{(1)}$.  Taking now into account the relation
(\ref{eq20}) between $f^{(0)}$ and $u_z^{(0)}$ at the contact
plane, as well as the expression (\ref{eq20a}) for $u_z^{(0)}$
there, we find for $\tilde{u}_z^{(1)}$ at $z=0$:
\begin{equation}
\label{eq:uztil}
\tilde{u}_z^{(1)}=\frac{(1-\nu_1^2)}{\pi Y_1} \int\!\!\!\int_S
\frac{ \beta_1 \dot{\sigma}^{el(0)}_{zz}(x',y', z=0)\,\mathrm{d}x'\mathrm{d}y'}{ \left|{\bf r} -{\bf r}'\right|}\,,
\end{equation}
where the subscript "1" indicates that the constant $\beta_1$ refers to the upper body.
Similar considerations may be done for the lower body, $z<0$, yielding:
\begin{eqnarray}
&&\hspace{-10mm}
\left.\bar{w}_z^{(1)}\right|_{z=0}=-\frac{(1-\nu_2^2)}{\pi Y_2}
\nonumber\\
\label{eq:wzbar}
&& \times\int\!\!\!\int_S
\frac{ \bar{\sigma}_{zz}^{el(1)}(\bar{\bf w}^{(1)}(x',y', z=0))\,\mathrm{d}x'\mathrm{d}y'}{ \left|{\bf r} -{\bf r}'\right|}\,,
\end{eqnarray}
and
\begin{equation}
\label{eq:wztil}
\tilde{w}_z^{(1)}=\frac{(1-\nu_2^2)}{\pi Y_2} \int\!\!\!\int_S
\frac{ \beta_2 \dot{\sigma}^{el(0)}_{zz}(x',y', z=0)\,\mathrm{d}x'\mathrm{d}y'}{ \left|{\bf r} -{\bf r}'\right|}\,.
\end{equation}
Now we apply the  consistency  condition (\ref{eq:consists}),
using Eq.~(\ref{eq19}) for the both bodies,
\begin{eqnarray}
&\hspace{-10mm}
\left. \left( \alpha_1 \dot{\sigma}_{zz}^{el(0)} + \bar{\sigma}_{zz}^{el(1)}(\bar{\bf u}^{(1)}) \right) \right|_{z=0}
\qquad\qquad &\nonumber\\[5pt]
\label{eq:cons2}
&\qquad\qquad=
\left. \left( \alpha_2 \dot{\sigma}_{zz}^{el(0)} + \bar{\sigma}_{zz}^{el(1)}(\bar{\bf w}^{(1)})\right) \right|_{z=0}\,,
&
\end{eqnarray}
where we also take into account that the following parts of the
stress tensor vanish on the contact plane:
$$
\left.
\tilde{\sigma}_{zz}^{el(1)}(\tilde{\bf u}^{(1)}) \right|_{z=0}
=\left.\tilde{\sigma}_{zz}^{el(1)}(\tilde{\bf w}^{(1)})
\right|_{z=0}=0.
$$
Eq.~(\ref{eq:cons2}) then yields,
\begin{eqnarray}
&&\hspace{-5mm}
\left. \bar{\sigma}_{zz}^{el(1)}(\bar{\bf w}^{(1)}) \right|_{z=0}
\nonumber\\
\label{eq:cons3}
 &&=
(\alpha_1  - \alpha_2  ) \left. \dot{\sigma}_{zz}^{el(0)}\right|_{z=0} +\left. \bar{\sigma}_{zz}^{el(1)}(\bar{\bf u}^{(1)})  \right|_{z=0}\,.
\end{eqnarray}
Now we use the boundary condition (\ref{eq:6}),
$$
 u_z^{(1)} + w_z^{(1)}=\bar{u}_z^{(1)} + \tilde{u}_z^{(1)} + \bar{w}_z^{(1)} + \tilde{w}_z^{(1)}=0\,,
$$
and applying  Eqs.~(\ref{eq:29b}), (\ref{eq:uztil}), (\ref{eq:6})
and (\ref{eq:wztil}) for $\bar{u}_z^{(1)}$, $\tilde{u}_z^{(1)}$,
$\bar{w}_z^{(1)}$ and $\tilde{w}_z^{(1)}$ we obtain,
\begin{eqnarray}
\int\!\!\!\int_S
\Big(( \beta_1 D_1 + \beta_2 D_2) \dot{\sigma}^{el(0)}_{zz}
 -D_1 \bar{\sigma}_{zz}^{el(1)}(\bar{\bf u}^{(1)})\qquad
\nonumber\\
\left. {} -D_2 \bar{\sigma}_{zz}^{el(1)}(\bar{\bf w}^{(1)}) \Big) \right|_{z=0}
\frac{ \mathrm{d}x'\mathrm{d}y'}{ \left|{\bf r} -{\bf r}'\right|}=0\,,
\nonumber
\end{eqnarray}
where we introduce the short-hand notations, $$D_i=
(1-\nu_i^2)/Y_i, \quad \qquad i=1,2.$$ From the above equation, together with
Eq.~(\ref{eq:cons3}) follows the relation for the first-order
elastic tensor:
\begin{eqnarray}
&&\left. \bar{\sigma}_{zz}^{el(1)}(\bar{\bf u}^{(1)})\right|_{z=0} =
 \left( \frac{ \beta_1 D_1 + \beta_2 D_2}{D_1+D_2}\right.
\\
\label{eq:sigmaelvis}
&& \left.\left. ~~~~~~~~~~~~~~~~~~~~-\frac{D_2 (\alpha_1 -\alpha_2 )}{D_1+D_2} \right) \dot{\sigma}^{el(0)}_{zz}\right|_{z=0}\,. \nonumber
\end{eqnarray}
Finally we obtain, taking into account that the total first-order
stress on the contact plane is a sum of two parts -- the elastic
one, given by Eq.~(\ref{eq:sigmaelvis}), and the dissipative part
from Eq.~(\ref{eq19}),
\begin{equation}
\label{eq:final_stress}
\left. \sigma_{zz}^{(1)}\right|_{z=0} = \left. ( \sigma_{zz}^{v(1)} +\sigma_{zz}^{el(1)} ) \right|_{z=0} =A  \left. \dot{\sigma}^{el(0)}_{zz}\right|_{z=0},
\end{equation}
where
\begin{equation}
\label{eq:A_def}
A= \frac{(\alpha_1+ \beta_1)D_1 + (\alpha_2 +\beta_2)D_2}{D_1+D_2}\,.
\end{equation}
Again  we take into account that the component
$\tilde{\sigma}_{zz}^{el(1)}(\tilde{\bf u}^{(1)})$ of the stress
tensor vanishes on the contact plane. The constant $A$ may be
written, using Eq.~(\ref{eq:alpha0}) and (\ref{eq28al}) for
$\alpha_{0(i)}$ and  $\alpha_{1(i)}$ as
\begin{eqnarray}
\label{eq:Afin}
A&=&\frac{\gamma_1D_1+\gamma_2D_2}{D_1+D_2} \\
\gamma_i&=&\left(\frac{1+\nu_i}{Y_i} \right)^2 \left[ \frac43 \eta_{1(i)}(1-\nu_i +\nu_i^2) + \eta_{2(i)} (1-2 \nu_i)^2 \right] \nonumber
\end{eqnarray}
Using the above Eqs.~(\ref{Pressure}), (\ref{Fel_xi_gen1}) and
(\ref{eq:xi_a}) we can write the explicit expression for the
viscous pressure
$P_z^{v(1)}=-\left.\sigma_{zz}^{(1)}\right|_{z=0}$ acting between
the colliding bodies:
\begin{equation}
\label{eq:Pzvis}
 P_z^{v(1)}(x,y) = -\frac{3A}{4 DN(\zeta_0)}\, \frac{\dot{\xi}}{\sqrt{a^2 -(x^2 +y^2 \zeta_0)}}\,,
\end{equation}
where $a$ depends on $\xi$ according to Eq.~(\ref{eq:xi_a}) and
all other notations have been introduced in the previous section.

\section{Dissipative Force}
Now we can write the dissipative force acting between particles.
It corresponds to the force associated with the viscous constants,
that is, with the first-order stress tensor $\sigma_{zz}^{(1)}$.
Integrating this stress over the contact area, we obtain,
$$
 F_z^{v(1)}=\int \!\!\!\int_S \sigma_{zz}^{(1)}(x,y)|_{z=0} \mathrm{d}x \mathrm{d}y \, ,
$$
so that Eq.~(\ref{eq:final_stress}) yields,
\begin{equation}
\label{eq:31a}
F_z^{v(1)} = -A \dot{F}_z^{el(0)},
\end{equation}
where $F_z^{el(0)}$ is the normal force corresponding to the
elastic reaction of the medium. It is equal to the Hertzian force,
Eq.~(\ref{Fel_xi_gen1}); taking the time derivative of this force
we finally obtain:
\begin{equation}
\label{eq:31b}
F_z^{v(1)} = -\frac 32 A C_0  \sqrt{\xi} \dot{\xi}\,.
\label{eq:31c}
\end{equation}
Here the constant $C_0$, defined by Eq.~(\ref{Fel_xi_gen1}), is
determined by the geometry of the colliding bodies and their
material properties (see the discussion after
Eq.~(\ref{Fel_xi_gen1})).

Hence the total force acting between two viscoelastic bodies reads
in the linear approximation with respect to the dissipative
constants:
\begin{equation}
\label{eq:31d}
F_{\rm tot} = C_0\xi^{3/2} -\frac 32 A C_0  \sqrt{\xi} \dot{\xi}\,,
\end{equation}
where the relation between the deformation $\xi$ and the axis $a$
of the contact ellipse is given by Eq.~(\ref{eq:xi_a}) as in the
static Hertz theory. Note however, that contrary to the Hertz
theory the size of the contact ellipse is determined now not by
the total force acting between the bodies, but by the elastic part of the total force,
$F_{\rm tot}+ (3/2)A C_0 \sqrt{\xi} \dot{\xi}$, that is, by apparently larger force for the compressive part of the impact ($\dot{\xi}>0$) and apparently smaller for the restoring part ($\dot{\xi}<0$).

\section{Conclusion}
We derive a new   expression for the dissipative force acting between viscoelastic bodies during an impact. Contrary to the previous theories, based on the physically plausible but non-rigorous approach, our theory exploits mathematically rigorous perturbation scheme with the small parameter being the ratio of the microscopic relaxation time and the impact duration. We make calculations for the zero and first-order terms in this perturbation expansion.  The new expression  for the dissipative force noticeably differs from the previous one, obtained within the quasi-static approximation. Due to rigorous derivation from the first principles we get a physically correct  result for the dissipative force acting between bodies of different materials; this was not possible within the previous approach. Moreover, our new theory is also lacking inconsistency of the previous theory with respect to materials with vanishingly small elastic shear module. While the previous, quasi-static theory predicts the nonphysical zero dissipation, the new theory implies dissipation, similar to that for "common" materials.

In the present study we neglect the inertial effects, that is, we assume that the characteristic velocity of the problem is much smaller than the speed of sound in the bodies.
The general approach presented in our study may be, however, further developed to take into account the inertial effects as well as high-order terms in the perturbation series.

\begin{acknowledgement}
DSG and AVP acknowledge financial support from the Russian Scientific Foundation (grant no.\ 14-21-00090).
\end{acknowledgement}


\end{document}